\documentstyle[12pt,fleqn]{article}
\input psfig

\newcommand{\bea}{\begin{eqnarray}}
\newcommand{\beq}{\begin{equation}}
\newcommand{\eea}{\end{eqnarray}}
\newcommand{\eeq}{\end{equation}}
\newcommand{\nnu}{\nonumber}

\newcommand{\lsim}{\;\raisebox{-0.9ex}{$\textstyle\stackrel{\textstyle<}{\sim}$}
 \;}

\begin{document}

\begin{titlepage}
\begin{flushright}
HEPHY-PUB 624/95\\
UWThPh-1995-9\\
June 1995
\end{flushright}
\vfill
\begin{center}
{\Large\bf $CP$--violating asymmetries in top--quark\\
production and decay in  $e^+e^-$ annihilation\\
\vspace{3mm}
within the MSSM}
\vfill
{\large Alfred Bartl}\\
{\em Institut f\"ur Theoretische Physik, Universit\"at Wien}\\
{\em A--1090 Vienna, Austria}
\vfill
{\large Ekaterina Christova} \\
{\em Institute of
Nuclear Research and
 Nuclear Energy}\\
{\em Boul. Tzarigradsko
Chaussee 72, Sofia 1784, Bulgaria.}
\vfill
{\large Walter Majerotto} \\
{\em Institut f\"ur Hochenergiephysik, \"Osterreichische
Akademie der Wissenschaften}\\
{\em A--1050 Vienna, Austria}
\vfill

{\sc Abstract}
\end{center} We obtain analytic formulae for the cross section of the
sequential processes of
$e^+e^-\rightarrow t\bar t$ and $t\rightarrow bl^+\nu$ / $\bar t \rightarrow
\bar b l^-\bar{\nu}$ in
the laboratory frame, where the  dependence on
 triple product correlations of the type $(\hat{\bf q}_1\times\hat{\bf
q}_2\cdot \hat{\bf q}_3)$,
induced by $CP$ violation both in the production and the decay are explicitely
shown. Different
observables sensitive to $CP$ violation are defined and calculated in the
Minimal Supersymmetric
Standard Model (MSSM). The observables sensitive to $CP$ violation are of the
order of $10^{-3}$. The
dependence on the masses of the supersymmetric particles is also shown.

\end{titlepage}

\newpage
\setcounter{footnote}{0}
\setcounter{page}{1}
\subsection*{1. Introduction}

Precision measurements of various production and decay modes of the
 top quark are expected to provide also information about
physics beyond the Standard Model (SM). Testing new physics
in observables which are sensitive to $CP$ violation seems
especially promising. As the top quark  does not
mix with other generations, the GIM mechanism of unitarity constraints
leads to negligibly small effects of $CP$  violation in the SM. Thus,
observation of $CP$  noninvariance in top--quark physics would
definitely be a signal for physics beyond the
 SM ~\cite{BernNP,BNM,ECMF1,ECMF2,Grzad}.

Here we shall consider $CP$ violation induced by supersymmetry (SUSY).
 In the Minimal Supersymmetric Standard Model (MSSM)
{}~\cite{Kane}, there are more possibilities to introduce complex
couplings than in the SM. Even without generation mixing,
 nonzero complex phases can occur in the Lagrangian
that cannot be rotated away by a
suitable redifinition of the fields. They give rise to $CP$
violation within a single generation, free of the unitarity supression of
the GIM mechanism.

In top--quark production and decay $CP$
violation is due to  radiative corrections. The
magnitude is determined by the ratio of the masses of the top quark
and the  SUSY particles in the
loop. Having in mind the large top--quark mass, $m_t=175 GeV$~\cite{??},
we may expect that the $CP$--violating effects are only
moderately suppressed. Thus, testing SUSY through $CP$--violating
observables in top--quark physics
is a promising task for future colliders.

Here, as possible evidence of $CP$ violation we consider $T$--odd triple
product
 correlations of the type
\beq
({\bf q_1q_2q_3}) \equiv ({\bf q_1 \times q_2 \cdot q_3})\, \label{triple}
\eeq
where ${\bf q_{1,2,3}}$ can be any one of  the
3--momenta  in $e^+e^-\rightarrow t\bar{t}$ or of the $t$ ($\bar{t}$)--
decay products $t\rightarrow bl^+\nu$  ($\bar{t}\rightarrow
\bar{b}l^-\bar{\nu}$). This method has been proposed
in~\cite{BernNP,BNM} for a general study of $CP$ violation in $t\bar
t$ production in $e^+e^-$ annihilation and in $pp$ collisions.
The correlations (1) are called $T$--odd as they
change sign under a flip of the 3--momenta involved. However, this does
neither imply time--reversal noninvariance nor $CP$ violation if
$CPT$ is assumed. When
loop corrections are included,  $T$--odd correlations can arise
either from absorptive parts in the amplitude (so-called
 final state interactions~\cite{Henley}), or from $CP$ violation.
 The former effect is a consequence of the unitarity of the $S$--matrix,
and it
can be eliminated either by taking the difference between the process we
are interested in and its $CP$ conjugate~\cite{Golowich} or by
direct estimates. $T$--odd correlations in the SM due to gluon
or Higgs boson exchange
in the final states have recently been considered in~\cite{NP}.

Note that in collider experiments  top--antitop quark pairs will be
copiously produced and the decay modes $t\rightarrow bl^+\nu$ and
$\bar{t}\rightarrow \bar{b}l^-\bar{\nu}$ will occur simultaneously.
Therefore, it will be possible to form the difference between the two
conjugate processes in the same experiment.

The appearance of triple product correlations of the type (1) can also
be explained by the fact that the top--quark polarization has
 non--vanishing transverse components both in the production
plane and  perpendicular to it.
Because of the large mass of the top quark the produced $t$ and
$\bar{t}$ can be regarded as free quarks with definite momenta and
polarizations that are not affected by hadronization. Thus their
polarization state can be infered from the distribution of their
decay products.

  In the $t\bar t$ production process
\beq
e^+(q_{\bar{l}}) + e^-(q_l )  \rightarrow t(p_t ) +
\bar{t}(p_{\bar{t}}) \, ,\label{production}
\eeq
a $T$--odd  correlation in the c.m.system is
$({\bf\xi q_lp_t})$~\cite{ECMF1}, where $\xi$ is the top--quark
polarization vector. Evidently, this correlation is different from
zero only if ${\bf\xi}$ has a component
 {\em normal to the production plane}.

In the semileptonic $t$ and $\bar{t}$  decays:
\bea
&&t\rightarrow b(p_b )l^+(p_{l^+})\nu (p_\nu )\, \label{tdecay}\\
&&\bar{t}\rightarrow \bar{b}(p_{\bar{b}})l^-(p_{l^-})\bar{\nu}
 (p_{\bar{\nu}} )\, \label{tbardecay}
\eea
a triple product correlation in the rest frame of the top quark
is  $({\bf\xi p_bp_{l^+}})$~\cite{ECMF2}. Here it is the polarization of
the top quark {\em normal to the decay plane} that contributes. In the
production as well as in the decay we shall only consider
 correlations due to $CP$ violation.

As the polarization of the top quark is not a directly observable
quantity, we obtain information about the above spin--momenta
correlations by triple product correlations of type (\ref{triple})
among the 3-momenta of the particles in the sequential processes:
\bea
e^+ &+& e^-  \rightarrow t + \bar{t} \,\rightarrow
\bar{t}\,b\,l^+\,\nu \label{seq}\\
e^+ &+& e^-  \rightarrow t +
\bar{t} \,\rightarrow t\,\bar{b}\,l^-\,\bar{\nu} .
\label{seq1}
\eea

In this paper we shall calculate the contribution to
 different triple product
correlations of  type (\ref{triple})
due to $CP$  violation in (\ref{production}) and in
 (\ref{tdecay}) or (\ref{tbardecay}). We shall obtain analytic
expressions for the sequential reactions (\ref{seq}) and
(\ref{seq1}) in which the triple product
correlations (\ref{triple}), sensitive to $CP$ violation in the
production and in the decay, explicitely appear. The problem of distinguishing
$
CP$ violation in
$t\bar{t}$  production and $t$ decay was considered previously in
{}~\cite{BernNP}
{}.
 We shall use the triple product correlations to investigate
$CP$ violation induced by SUSY. However, our method is
completely general and can also be applied to study other
sources of $CP$ violation.

In section 2 we obtain general expressions of processes
(\ref{seq}) and (\ref{seq1})
 using the formalism developed in ~\cite{BG}.
We write them in a form in which $CP$--violation manifests
itself in triple product correlations.
In section 3 we obtain the expressions for
the polarization vectors of the top and antitop quarks, when the
electric and weak dipole--moment couplings are taken into
account. Formulae for the cross section of (\ref{seq}) and
(\ref{seq1}) in the c.m.system with the explicit dependence on the triple
products of type (\ref{triple}) are given in section 4.
In section 5 we formulate the $CP$ -- violating observables.
 In section 6 we obtain
numerical results for these effects in the MSSM.

\subsection*{2. The Cross Section}

  Here we obtain  analytic expression for the cross
section of the sequential processes (\ref{seq}) and (\ref{seq1}),
in terms of the top/antitop polarization vectors.
 CP violation in both the production
and the $t(\bar{t})$-decay processes are taken into account.

We write the amplitudes of (\ref{seq}) and (\ref{seq1}) in the form:
\bea
{\cal M}^{t,\bar{t}} & =&  \frac{g}{2\sqrt{2}}\: \bar{u}
(p_\nu)\gamma_\alpha (1\mp \gamma_5)u(-p_{l^+}) \frac
{-g^{\alpha\beta}+\frac{p_W^\alpha\,p_W^\beta}{m^2_W}}
{p_W^2-m_W^2+im_W\Gamma
_W} \nnu \\ & &\times \bar{u}(p_b)V_\beta^{t,\bar{t}} \frac
{\rlap/p_t + m_t}{p_t^2-m_t^2+im_t\Gamma _t}\,
A^{t,\bar{t}}\,u(-p_{\bar{t}})\,.\label{M}
\eea

 Here $A^{t,\bar{t}}$ enters the amplitude $M$ for the production process
(\ref{production}) which we write in the two equivalent forms:
\beq
M =\bar{u}(p_t)A^t u(-p_{\bar{t}})
=\bar{u}_c(p_{\bar{t}})A^{\bar{t}} u_c(-p_t)\,. \label{M1}
\eeq
$A^t$ and $A^{\bar{t}}$ are given by s--channel $\gamma$ and
$Z^0$ exchange,
\bea
A^{t,\bar{t}}& =&  i\frac {e^2}{s}\:\bar{u}(-q_{\bar{l}})\,
\gamma_\alpha \,u(q_l) ({\cal V}_\gamma^{t,\bar{t}})^\alpha \nnu\\
&&-i\frac{g_Z^2}{s-m_Z^2}\:\bar{u}(-q_{\bar{l}})\,
\gamma_\alpha (c_V+c_A\gamma_5)\,u(q_l)({\cal V}_Z^{t,\bar{t}})^\alpha
,\label{A1}
\eea
where
$c_V=-1/2+2\sin^2\theta_W\,, c_A=1/2$ are
 the SM couplings of $Z^0$ to the
electron, and $\sqrt{s}$ is the total c.m.energy. The quantities
 $({\cal V}_i^{t,\bar{t}})$ describe the
$t\bar{t}\gamma$ and $t\bar{t}Z^0$ vertices:
\bea
({\cal V}_\gamma^{t,\bar{t}})_\alpha & =& \frac{2}{3}
\gamma_\alpha \mp\frac{d^\gamma}{m_t} {\cal P}_\alpha \gamma_5,\nnu\\
({\cal V}_Z^{t,\bar{t}})_\alpha & =& \gamma_\alpha (g_V \pm
g_A\gamma_5 )\mp\frac{d^Z}{m_t} {\cal P}_\alpha \gamma_5,
\label{prodvertex}
\eea
with  ${\cal P} = p_t-p_{\bar{t}}$. $g_V=1/2-4/3\,\sin^2\theta_W$
and $g_A=-1/2$  in eq. (10)  are the
SM top--quark couplings to the $Z^0$, $ d^\gamma$ and $d^Z$
are the electric and
weak dipole moments of the $t$-quark, which in the limit of
vanishing electron mass present the whole effect of $CP$ violation.
 They can be induced only by $CP$--violating interaction and have in general
  both real and imaginary parts.

The $t(\bar{t})$--decays (\ref{tdecay}),  (\ref{tbardecay})
are treated as a sequence of the two--body
decays $t\,\rightarrow \,b W^+(p_W)$ and $W^+\,\rightarrow
\,l^+\,\nu$, ($\bar{t}\,\rightarrow
\,\bar{b}\,
W^-(p_W)$ and $W^-\,\rightarrow
\,l^-\bar{\nu}$ ).  We write the $tbW$ and $\bar t\bar b W$
vertices in the form:
\bea
V^{t}_\alpha & =& \frac{g}{2\sqrt{2}}\left(\gamma_\alpha (1-\gamma_5) +
 f_L^t\gamma_\alpha (1-\gamma_5 ) + \frac{g_R^{t}}{m_W}
P^{t}_\alpha (1+ \gamma_5 )\right) ,\label{decayvertex} \\
V^{\bar{t}}_\alpha &=& \frac{g}{2\sqrt{2}}\left(\gamma_\alpha
(1+\gamma_5) + f_L^{\bar{t}*}\gamma_\alpha (1+\gamma_5 ) +
\frac{g_R^{\bar{t}*}}{m_W} {P}^{\bar{t}}_\alpha (1- \gamma_5
\right) \label{decayvertex1}
\eea
with $g$ the weak coupling constant and ${P}^t = p_t+p_b$, ${P}^{\bar{t}} =
p_{\bar{t}}+p_{\bar{b}}$.  In eqs. (\ref{decayvertex}) and
(\ref{decayvertex1}) we have kept only the terms that do not vanish
in the approximation $m_b = 0$. The form factors
$f_L^{t,\bar{t}} $ and $g_R^{t,\bar{t}}$ get contributions both from
the $CP$--invariant absorptive parts of the amplitudes and from
$CP$--violating interactions, and can have real and imaginary parts.
Furthermore, $CPT$ is assumed. Neglecting the absorptive parts we
 then have ~\cite{BernNP}:
\beq
f_L^t=\left(f_L^{\bar{t}}\right)^*\equiv f_L,\qquad
g_R^t=\left(g_R^{\bar{t}}\right)^*\equiv g_R
\eeq
and $CP$ invariance  implies
\beq
f_L=f_L^*,\qquad g_R=g_R^*
\eeq

For the $Wl\nu$--vertices we take the SM tree level expressions.

The amplitude ${\cal M}^{\bar{t}}$ for process (\ref{seq1}) is obtained
from (\ref{M}) when also
the spinors
$\bar{u}(p_\nu)$, $u(-p_{l^+})$,  $\bar{u}(p_b)$ and
$u(-p_{\bar{t}})$ are replaced  by
$\bar{u}_c (p_{\bar{\nu}})$, $u_c(-p_{l^-})$,
$\bar{u}_c(p_{\bar{b}})$ and  $u_c(-p_t)$, respectively,
 which describe the charge conjugate particles.

In order to obtain the general expression for the cross sections of
(\ref{seq}) and (\ref{seq1}) we use the
formalism developed in~\cite{BG}.

In the narrow width
approximation for the top and the $W$ ($\Gamma_t \ll m_t , \Gamma_W
\ll m_W$ ) we obtain the cross section of reaction (\ref{seq}) in
 the form:
\beq
 d\sigma =
 d\sigma_{e^+e^-} \cdot d\Gamma_{\vec {t}}\cdot
\frac {p_{t_0}}{m_t \Gamma _t}\cdot d\Gamma_{\vec{W}}
\frac {p_{W_0}}{m_W \Gamma_W}\,.\label{sigma}
\eeq

Here $ d\sigma_{e^+e^-}$ is the differential cross section for $t
\bar{t}$ production (\ref{production}) in the c.m.s., $d\Gamma_{\vec
{t}}$ is the decay rate for $t \rightarrow b W$, with the
polarization state of the top quark determined in the production
process $e^+e^- \rightarrow t\bar{t}$, and
$d\Gamma_{\vec {W}}$ is the partial decay rate of $W \rightarrow
 l\nu $, with the polarization state of the $W$ determined in the
preceding $t$ decay.  $d\Gamma_{\vec {t}}$ and $d\Gamma_{\vec {W}}$ are
the decay distributions in the c.m.s. of the initial $e^+ e^-$. As $
m_t\Gamma_t/p_{t_0}\:\left( m_W\Gamma_W/p_{W_0}\right)$ is the total
decay width of $t\:\left(W\right)$ in a frame in which the momentum
of the top ($W$) equals ${\bf p}_t\:\left({\bf p}_W\right)$,
$d\Gamma_{\vec {t}}\,\cdot p_{t_0}/m_t\Gamma_t\:
\left(d\Gamma_{\vec {W}}\,\cdot p_{W_0}/m_W\Gamma_W\right)$
is the branching ratio of the decay  $t\rightarrow
b\,W\:\left(W\rightarrow l\nu \right)$ in the laboratory frame, with
the polarization of $t(W)$ determined in the preceding process.

 Now we are interested only in triple product correlations in reactions
(\ref{seq}) and (\ref{seq1}) induced
by $CP$ violation in ${\cal V}_\alpha^i$ and/or $V_\alpha$. First we
consider process (\ref{seq}).
Possible triple product correlations are
$({\bf q}_l{\bf p}_t{\bf p}_b)$,
$({\bf q}_l{\bf p}_t{\bf p}_{l^+})$,
$({\bf q}_l{\bf p}_{l^+}{\bf p}_b)$, and
$({\bf p}_t{\bf p}_{l^+}{\bf p}_b)$.  These
correlations follow from the covariant quantities $\varepsilon (p_1p_2p_3p_4)$,
where
$p_i$ can be any one of the 4--vectors
$q_l,\,q_{\bar{l}},\,p_t,\,p_b,\,p_{l^+}$ , when  written in the laboratory
frame. (The symbol
$\varepsilon (p_1p_2p_3p_4)$ means $\varepsilon_{\alpha\beta\gamma\delta}\,
p_1^\alpha\,p_2^\beta\,p_3^\gamma\,p_4^\delta$ with $\varepsilon_{0123}=-1$.)

 The  top--quark polarization normal to the production plane,
$\xi_N$, which is induced by the $CP$--violating dipole couplings,
gives rise to $T$--odd triple product correlations
in $e^+ e^-\rightarrow t\bar{t}$ which are proportional to $\Im m d^i$
\cite{ECMF1}. CP--violating triple product correlations in the decay
$t\rightarrow b\,l^+\,\nu$ arise from the top polarization {\it
transverse to the decay plane} and are induced by $\Im m\,
g_R$ in the $tbW$--vertex
{}~\cite{ECMF2}. Both types of couplings $d^i$ and $g_R$ are
generated by loop corrections.

We write the polarization 4--vector $\xi^t$ and
 $\xi^{\bar t}$ as the sum
\beq
\xi^{t,\bar{t}} =\xi^{t,\bar{t}}_{SM} + \xi^{t,\bar{t}}_N,
\eeq
where $\xi_{SM}$ is the tree level SM contribution that  lies in the
production plane, and $\xi_N$  is the component normal to the
production plane, which can arise in general either from
final-state interactions or from $CP$--violating interactions.
Neglecting $\Re e g_R$ we obtain the following expression for the cross
sections of (\ref{seq}) and (\ref{seq1}) in terms of the
top/antitop quark polarization vectors:
\beq
d\sigma^{t,\bar t} = \frac{96(4\pi )^2}{s}\alpha^2\left(\frac{g} {2\sqrt
2}\right)^4
\left( m^2_t-2(p_tp_{l^+})\right)N\left\{ A_{SM}^{t,\bar t}
+ A_d^{t,\bar t} +
A_{g_R}^{t,\bar t}\right\}\:d\Gamma^{t,\bar t} \label{sigmat}
\eeq

Here $\alpha $ is the finestructure constant and
\bea
A_{SM}^{t,\bar t} & =& (p_tp_{l^+})\mp m_t(\xi_{SM}^{t,\bar t}
 p_{l^+})\, ,\label{ASM}\\
A_d^{t,\bar t} & = & \mp m_t\,(\xi^{t,\bar t}_N p_{l^+})\,, \label{AD}\\
 A_{g_R}^{t,\bar t} & = & \mp 2\frac{\Im m g_R}{m_W} \varepsilon
(\xi^{t,\bar t}_{SM} p_t p_{l^+} p_b)\,, \label{AGR}\\
N &=& \left( 1+\beta^2\cos^2\theta \right) F_1
+\frac{4m_t^2}{s}F_2+2\beta\cos\theta F_3\label{N}\\
F_1 &=& \left(\frac{2}{3}\right)^2 +
h_Z^2(c_V^2+c_A^2)(g_V^2+g_A^2)-\frac{4}{3}h_Zc_Vg_V\nnu\\
F_2 &=& \left(\frac{2}{3}\right)^2 +
h_Z^2(c_V^2+c_A^2)(g_V^2-g_A^2)-\frac{4}{3}h_Zc_Vg_V\nnu\\
F_3&=& 4 h_Z c_Ag_A\left(h_Zc_Vg_V-\frac{1}{3}\right)
\eea

 $A_{SM}^{t,\bar{t}}$
is the SM contribution with
the first terms describing the contribution where the produced
top and antitop are unpolarized, while the second one  takes into
account $\xi^{t,\bar{t}}_{SM} $.

The terms $A_d^{t,\bar{t}}$ and $A_{g_R}^{t,\bar{t}}$ contain the $T$--odd
$CP$--violating
correlations induced by $\Im m d^{\gamma}$ and $\Im m d^Z$ via
$\xi^{t,\bar{t}}_N$, and by $\Im m
g_R$, respectively. (Note that the form factor $f_L$ of eqs.\ref{decayvertex}
and
\ref{decayvertex1} does not appear.) The complete expressions for
$\xi^{t,\bar{t}}$ will be given in
the next section. The invariant phase space element $d\Gamma $ is \bea
d\Gamma^t & =& \frac{1}{(2\pi
)^8} \delta (q_l+q_{\bar{l}}-p_{l^+} -p_\nu -p_b-p_{\bar{t}})\cdot \frac{\pi
}{m_t\Gamma_t}\delta
(p_t^2-m_t^2)\nnu\\
& &\times \frac{\pi}{m_W\Gamma_W} \delta(p_W^2-m_W^2)
\frac{d{\bf p}_{l^+}}{2p_{l\,0}}\, \frac{d{\bf
p}_\nu}{2p_{\nu\,0}}\,\frac{d{\bf
p}_b}{2p_{b\,0}}\,\frac{d{\bf p}_{\bar{t}}}
{2p_{\bar{t}\,0}}\,.\label{Gamma}
\eea
Here  $\beta = \sqrt {1-4m_t^2/s}$ is
 the velocity factor of the $t$--quark in the c.m.system,
 $\theta$ is the angle between the momenta of the initial electron
and the $t$--quark, and

\beq
h_Z=\frac{s}{s-m_Z^2}\cdot\frac{g_Z^2}{e^2}\,,\qquad
g_Z=\frac{e}{\sin 2\theta_W}\,,\\
\eeq

In order to obtain the cross section for reaction (\ref{seq1}), in the
expressions (\ref{sigmat}) -- (\ref{AGR}) and(\ref{Gamma})
 the following replacements must be made:
\beq
p_t\rightarrow p_{\bar t},\qquad p_b\rightarrow p_{\bar b}, \qquad
p_{l^+}\rightarrow p_{l^-},\qquad p_\nu\rightarrow
 p_{\bar{\nu}}\,.\label{repl}
\eeq

\subsection*{3. The top--quark polarization vector}

First we calculate the polarization vector $\xi_\alpha^{t} $
of the top quark. It is determined in the production process,
and it is given by the expression~\cite{BG}:
\bea
\xi_\alpha^t &= &
\left( g_{\alpha\beta} - \frac{p_{t\alpha}p_{t\beta}}{m^2_t}\right)
\cdot Tr\:
\left( A^t (-\Lambda (-p_{\bar{t}}))\bar{A^t}\Lambda
(p_t)\gamma^\beta\gamma_5\right) \nnu \\ & &\times
\left\{ Tr\left[ A^t (-\Lambda
(-p_{\bar{t}}))\bar{A^t}\Lambda (p_t )\right]\right\}^{-1}\,.\label{xit}
\eea

$\xi_\alpha^t$ can be decomposed along three vectors which are
independent and orthogonal to $p_t$: two of them, $Q_l$ and
$Q_{\bar l}$
\beq
Q_l = q_l - \frac {p_tq_l}{m_t^2}\cdot p_t\qquad Q_{\bar{l}} =
q_{\bar{l}} - \frac {p_tq_{\bar{l}}}{m_t^2}\cdot p_t
\eeq
are in the production plane, and the third,
 $ \varepsilon^{\alpha\beta\gamma\delta} p_{t\beta}
q_{l\gamma} q_{\bar{l}\delta}$, is normal to it. We then have:
\beq
(\xi^t_{SM})^\alpha ={P}_l^t(Q_l)^\alpha +
{P}_{\bar{l}}^t(Q_{\bar{l}})^\alpha \qquad (\xi^t_
N)^\alpha =
{D}^t \varepsilon^{\alpha\beta\gamma\delta} p_{t\beta}
q_{l\gamma} q_{\bar{l}\delta} .\label{xi1}
\eeq

We neglect contributions of $\Im m d^\gamma$ and $\Im m d^Z$ to
$P_l^t$ and $P_{\bar{l}}^{t}$ as
they are much smaller than the SM contributions. Together with eqs.
(\ref{A1}) and  (\ref{prodvertex}),
 up to first order in  $\Im m d^\gamma$ and $\Im m d^Z$,
 eq. (\ref{xit}) implies the following expressions for
 the three components of the polarization vector
as defined above:
\bea
{P}_l^t &=&\frac{2m_t}{s}\left\{
\left(1-\beta\cos\theta\right)
\,G_1+\left(1+\beta\cos\theta\right)\right.\,G_2\nnu\\
& &\left.-\left(3-\beta\cos\theta\right)\,G_3\right\}/N\,,\label{Pl}\\
{P}_{\bar{l}}^t
&=&-\frac{2m_t}{s}\left\{\left(
1+\beta\cos\theta\right)
\,G_1+\left(1-\beta\cos\theta\right)\,G_2\right.\nnu\\
& &\left.+\left(3+\beta\cos\theta\right)\,G_3\right\}/N\,,\label{Pl1}\\
{D}^t &=&\frac{8}{sm_t}\,
\left\{D_1+\beta
\cos\theta D_2\right\}/N.\label{D}
\eea

Here
\bea
G_1&=&2h_Z^2\,c_Vc_A\left(g_V^2+g_A^2\right)-\frac{4}{3}
h_Z\,c_Ag_V\,,\\
G_2&=&2h_Z^2\,c_Vc_A\left(g_V^2-g_A^2\right)-\frac{4}{3}
h_Z\,c_Ag_V\,,\\
G_3&=&2h_Z^2\,\left(c_V^2+c_A^2\right)g_Vg_A-\frac{4}{3}
h_Z\,c_Vg_A\,,\\
D_1&=&2h_Z^2\, c_Vc_Ag_V\,\Im m\,d^Z
-h_Z\,c_A\left(g_V\Im m\,d^\gamma+\frac{2}{3}\Im
m\,d^Z\right)\,,\\
D_2&=&h_Z^2\,(c_V^2+c_A^2)g_A\,\Im
m\,d^Z-h_Z\,c_Vg_A\Im m\,d^\gamma
\eea

As can be seen from eqs. (\ref{xi1}),(\ref{D}),(36),(37), $\Im m d^\gamma$ and
$
\Im m d^Z$
enter only in the polarization component transverse to the production plane.

In an analogous way we write the polarization vector  $\xi_{SM}^{\bar{t}}$
 of the antitop quark as
\beq
(\xi_{SM}^{\bar{t}})^\alpha =
{P}_l^{\bar{t}}\bar{Q}_l^\alpha +
{P}_{\bar{l}}^{\bar{t}}\bar{Q}_{\bar{l}}^\alpha ,\qquad
(\xi_N^{\bar{t}})^\alpha ={D}^{\bar{t}}
\varepsilon^{\alpha\beta\gamma\delta} p_{\bar{t}\beta} q_{l\gamma}
q_{\bar{l}\delta}\,,\label{xitbar}
\eeq
where
\beq
\bar{Q}_{l\left(\bar{l}\right)} =Q_{l\left(\bar{l}\right)}
(p_t\rightarrow p_{\bar{t}})\,.
\eeq
Then we obtain
\beq
{P}_l^{\bar{t}}=-{P}_{\bar{l}}^{t},\qquad
{P}_{\bar{l}}^{\bar{t}}=-{P}_l^t ,\qquad
 D^{\bar{t}}=D^t \,.\label{antitop}
\eeq

In the c.m.system, the expressions (27) -- (31) and (37) -- (39)
for $\xi^{t,\bar{t}}$ lead to the following relations
between the space components of the polarization vectors of
the top and the antitop:
\bea
\vec{\xi}_{SM}^{\,\,\bar{t}}&=&\vec{\xi}_{SM}^{\,t}\label{xiSM}\\
\vec{\xi}_{N}^{\,\,\bar{t}}&=&-\vec{\xi}_{N}^{\,t}\,.\label{xiSUSY}
\eea

This result can be readily understood by applying the $CP$ operator
(and is also a check of our results): $CP$ invariance implies
$\vec{\xi}^t=\vec{\xi}^{\,\bar{t}}$ which is nothing but the result
 (\ref{xiSM}) for
the tree level SM (i. e. $CP$ conserving) contribution.
Evidently, eq. (\ref{xiSUSY}) violates $CP$, being induced by $d^\gamma$
 and $d^Z$.

Eq. (\ref{antitop}) implies:
\bea
A_{SM}^{\bar{t}}(p_{\bar{t}},p_{l^-},p_{\bar{b}})&=&
A_{SM}^t(p_t,p_{l^+},p_b),\\
A_{d}^{\bar{t}}(p_{\bar{t}},p_{l^-},p_{\bar{b}})&=&
-A_{d}^t(p_t,p_{l^+},p_b),\\
A_{g_R}^{\bar{t}}(p_{\bar{t}},p_{l^-},p_{\bar{b}})&=&
-A_{g_R}^t(p_t,p_{l^+},p_b).
\eea

\subsection*{4. Triple product correlations}

 Now we obtain the explicit dependence of the cross section of
processes (\ref{seq}) and (\ref{seq1}) on triple product
correlations of type (\ref{triple}) in the c.m.system, the
laboratory frame of the initial $e^+e^-$ beams.

 From
(\ref{sigmat}), (\ref{xit}) and (\ref{xitbar}) we obtain the cross section
$d\sigma^{t,\bar t}$ in
the c.m. system.
\bea
d\sigma^{t,\bar{t}} =
 \sigma_{SM}^{t,\bar{t}}\left\{1\right. &+& \frac{1}{1-\beta
  (\hat{{\bf p}}_t\hat{{\bf p}}_{l^+})}
 \left[\left(\hat{{\bf q}}_l\hat{{\bf
p}}_t\hat{{\bf p}}_{l^+}\right)A^{t,\bar{t}}_1 +
 \left(\hat{{\bf q}}_l\hat{{\bf p}}_t\hat{{\bf
p}}_b\right)A^{t,\bar{t}}_2\right.\nnu \\
& +& \left(\hat{{\bf p}}_t\hat{{\bf p}}_{l^+}
\hat{{\bf p}}_b\right)A^{t,\bar{t}}_3
+\left.\left.\left(\hat{\bf q}_l\hat{\bf p}_{l^+}\hat{\bf
p}_b\right)A^{t,\bar{t}}_4\right]\right\}d\Omega_t d\Omega_b d\Omega_l
 \label{sigma3}
\eea
where $\hat{q}_e , \hat{p}_t$, etc.  denote the corresponding unit 3-vectors.

We have
\beq
\sigma_{SM}^{t,\bar{t}}
 =\frac{\alpha^2}{4\pi^4}\left(\frac{g}{2\sqrt 2}\right)^4
\frac{\beta}{s}\frac{ \left[m_t^2-2(p_tp_{l^+})\right]}
{ m_t\Gamma_t\,m_W\Gamma_W}\frac{E_b^2}
{m_t^2-m_W^2}\frac{E_l^2}{m_W^2}N A_{SM}^{t,\bar t}
\eeq
which is the expression for the SM cross sections of (\ref{seq}) and
(\ref{seq1}). Here $E_b$ and $E_l$ are the energies of the
the b-quark and the final lepton in the c.m.s.
\beq
E_b=\frac{m_t^2-m_W^2}{2E}\frac{1}{1-\beta (\hat{{\bf p}}_t\hat{{\bf
p}}_b)}
\eeq
\beq
 E_l=\frac{m_W^2}{2\left[ E(1-\beta (\hat{{\bf p}}_t\hat{{\bf
p}}_{l^+}))-E_b(1-(\hat{{\bf p}}_b\hat{{\bf
p}}_{l^+}))\right]}
\eeq
and $E = \sqrt{s}/2$. For the asymmetries $A_i$ we obtain:
\bea
A^{t,\bar{t}}_1 &=&
\mp\beta\left[\frac{D}{2}-\frac{m_t}{\sqrt s}E_b {P}_-
\frac{\Im m g_R}{m_W}\right]\label{alpha1} \\
A^{t,\bar{t}}_2 &=& \mp\beta\frac{m_t}{\sqrt s} E_b P_
-\frac{\Im m g_R}{m_W}\label{alpha2} \\
A^{t,\bar{t}}_3 &=&- \
\beta\frac{m_t}{\sqrt s}E_b {P}_+\frac{\Im m g_R}{m_W}
\label{alpha3}\\
A^{t,\bar{t}}_4 &=& \pm\frac{m_t}
{\sqrt s} E_b {P}_- \frac{\Im m g_R}{m_W}
\label{alpha4}
\eea
where
\bea
P_-&=&\frac{4}{N}\left[G_1+G_2+\beta\cos\theta\, G_3\right]\nnu\\
P_+&=&-\frac{4}{N}\left[3G_3+\beta\cos\theta\, (G_1-G_2)\right]\nnu\\
D&=&\frac{8}{N}\left[D_1+\beta\cos\theta\, D_2\right]\,,
\eea
$N$ was defined in eq.(\ref{N}).

In order to obtain the expressions (\ref{sigma3}) --
(\ref{alpha4})  for the reaction (\ref{seq1})
(with the upper index  $\bar{t}$) also
the replacements
(\ref{repl}) have to be made.

Notice, that only the correlation
$(\hat{\bf q}_l\hat{\bf p}_t\hat{\bf p}_l)$ gets a CP violating contribution
from both the
production and the decay, whereas the other correlations are sensitive only to
CP violation in the
decay.  From eqs. (\ref{alpha1}) -- (\ref{alpha4}) one expects that the triple
product correlations
discussed above are less sensitive to $CP$ violation in top--decay then in
$t\bar t$ production,
because $\Im m g_R$ is multiplied by the SM--polarization $P_\pm$ and
by $E_b/\sqrt{s}$.

We want to remark that  $(\hat{\bf q}_l\hat{\bf p}_t\hat{\bf p}_l)$,
$(\hat{\bf q}_l\hat{\bf p}_t\hat{\bf
p}_b)$ and  $(\hat{\bf q}_l\hat{\bf p}_l\hat{\bf p}_b)$ do not change sign
under the replacements
(\ref{repl}) when calculating $d\sigma^{\bar{t}}$ in the c.m.s., while
$(\hat{\bf p}_t\hat{\bf p}_l\hat{\bf p}_b)$ changes sign.
Together with  eqs.
(\ref{alpha1})--(\ref{alpha4}) this implies that a nonzero value of
the difference of $d\sigma^t$
and $d\sigma^{\bar t}$ would be the  genuine  signal for
a triple product correlation induced by $CP$ violation.

\subsection*{5. Observables}

We consider two types of observables:

(i) If $N\left[(\hat{\bf q}_l\hat{\bf p}_t\hat{\bf p}_l)>0 (<0)\right]$
is the number of events in which
$ (\hat{\bf q}_l\hat{\bf p}_t\hat{\bf p}_l)>0 (<0)$, with
analogous definitions for the other triple products,
  we define the following $T$--odd asymmetries:
 \bea
 O_1& =& \frac{N\left[(\hat{\bf q}_l\hat{\bf p}_t\hat{\bf p}_l)>0\right]-
 N\left[(\hat{\bf q}_l\hat{\bf p}_t\hat{\bf p}_l)<0\right]}
 {N\left[(\hat{\bf q}_l\hat{\bf p}_t\hat{\bf p}_l)>0\right]+
 N\left[(\hat{\bf q}_l\hat{\bf p}_t\hat{\bf p}_l)<0\right]}\label{O1}\\
 O_2& =& \frac{N\left[(\hat{\bf q}_l\hat{\bf p}_t\hat{\bf p}_b)>0\right]-
 N\left[(\hat{\bf q}_l\hat{\bf p}_t\hat{\bf p}_b)<0\right]}
 {N\left[(\hat{\bf q}_l\hat{\bf p}_t\hat{\bf p}_b)>0\right]+
 N\left[(\hat{\bf q}_l\hat{\bf p}_t\hat{\bf p}_b)<0\right]}\label{O2}\\
 O_3& =& \frac{N\left[(\hat{\bf p}_t\hat{\bf p}_l\hat{\bf p}_b)>0\right]-
 N\left[(\hat{\bf p}_t\hat{\bf p}_l\hat{\bf p}_b)<0\right]}
 {N\left[(\hat{\bf p}_t\hat{\bf p}_l\hat{\bf p}_b)>0\right]+
 N\left[(\hat{\bf p}_t\hat{\bf p}_l\hat{\bf p}_b)<0\right]}\label{O3}\\
 O_4& =& \frac{N\left[(\hat{\bf q}_l\hat{\bf p}_l\hat{\bf p}_b)>0\right]-
 N\left[(\hat{\bf q}_l\hat{\bf p}_l\hat{\bf p}_b)<0\right]}
 {N\left[(\hat{\bf q}_l\hat{\bf p}_l\hat{\bf p}_b)>0\right]+
 N\left[(\hat{\bf q}_l\hat{\bf p}_l\hat{\bf p}_b)<0\right]}\label{O4}
 \eea

(ii) The other $T$--odd observables we
 consider are the mean values of the triple product
correlations:
\bea
M_1=\langle (\hat{\bf q}_l\hat{\bf p}_t\hat{\bf p}_l)\rangle\label{aean}\\
M_2=\langle (\hat{\bf q}_l\hat{\bf p}_t\hat{\bf p}_b)\rangle\label{bean}\\
M_3=\langle (\hat{\bf p}_t\hat{\bf p}_l\hat{\bf p}_b)\rangle\label{cean}\\
M_4=\langle (\hat{\bf q}_l\hat{\bf p}_l\hat{\bf p}_b)\rangle\label{dean}
\eea

 The truly $CP$--violating effect would be a nonzero value of
 the differences:
 \bea
{\cal O}_i &=& O_i-\bar{O}_i\label{calO}\\
{\cal M}_i &=& M_i-\bar{M}_i\label{calM}
 \eea
where $O_i$ and $M_i$ refer to (\ref{seq}) and $\bar{O}_i$  and $\bar{M}_i$
 refer to reaction (\ref{seq1}).

Some of these correlations (${\cal M}_1$ and ${\cal M}_4$)
have been considered previously in~\cite{BernNP}

\subsection*{6. Minimal Supersymmetric Standard Model}

 In this section we shall give numerical predictions  in the
MSSM~\cite{Kane} for the observables defined in the preceding
section as well as for the quantities $d^\gamma$, $d^Z$,
$\Im m g_R$, and $\xi_N^{t,\tilde{t}}$.

The SM  contribution is too small to be of any interest to
be measured. Due to flavour mixing of the three generations
of quarks $d^i$ is  a two--loop effect. The contribution to
$\Im m g_R$ has been estimated in ~\cite{Ma}.

In the MSSM  $d^\gamma$, $d^Z$ and $\Im m g_R$ are generated at one--loop
order, irrespectively of generation mixing.
The main contribution comes from
 diagrams with gluino and scalar quarks  in the loop, as shown in Fig. 1. The
$CP$--violating phases $\phi_A$ and $\phi_{\tilde g}$ appear in the stop--
squark mixing
matrix and in the Majorana mass term of the gluino, respectively ~\cite{Dugan}.

The dipole moments are determined by the  gluino -- stop exchange
loops in  the
$t\bar t \gamma$ and $t\bar t Z^0$ vertices (see Fig. 1) and are given by
the following expressions:
\bea
\frac{ d^\gamma}{m_t}&=&\frac{\alpha_s}{3\pi}\frac{2}{3}\tilde{m}_g\sin
2\tilde{\theta}\sin
(\phi_A-\phi_{\tilde{g}})\left[I_{11}-I_{22}-2\left(I_{11}^\prime
 -I_{22}^\prime \right)\right]\label{dgamma}\\
 \frac{ d^Z}{m_t}&=&\frac{\alpha_s}{3\pi}\tilde{m}_g\sin
2\tilde{\theta}\sin
(\phi_A-\phi_{\tilde{g}})\sum_n\left(\vert a_n^L\vert^2-
\frac{4}{3}\sin^2\theta_W\right)\nnu\\
&&\times\left[I_{1n}-I_{2n}-2\left(I_{1n}^\prime
 -I_{2n}^\prime \right)\right]\label{dZ}
\eea

 The $CP$--violating phases enter only in the combination
  $\phi_A-\phi_{\tilde{g}}$. $\tilde\theta$ is the stop --
mixing angle which transforms the stop
mass eigenstates $\tilde{t}_n$, $n=1,2$, with masses $m_{\tilde{t}_n}$,
to the weak eigenstates $\tilde{t}_L$ and $\tilde{t}_R$:
\bea
\tilde{t}_L& =& exp (-i\phi_A/2)\left(\cos\tilde\theta\,\tilde{t}_1
+\sin\tilde\theta\,\tilde{t}_2\right) \equiv a_n^L\cdot\tilde{t}_n\\
\tilde{t}_R& =& exp (i\phi_A/2)\left(-\sin\tilde\theta\,\tilde{t}_1
+\cos\tilde\theta\,\tilde{t}_2\right) \equiv a_n^R\cdot\tilde{t}_n\,.
\eea

We have also used the notation:
\bea
I_{n,m} & =& \int
\frac{d^4k}{\pi^2}\frac{1}{k^2-\tilde{m}_g^2}
\frac{1}{\left(p_t-k\right)^2-\tilde{m}_n^2}
\frac{1}{\left(p_{\bar t}+k\right)^2-\tilde{m}_m^2}\label{int1}\\
{\cal P}_\alpha I_{n,m}^\prime+{\cal Q}_\alpha I_{n,m}^{\prime\prime}
& =& \int
\frac{d^4k}{\pi^2}k_\alpha\frac{1}{k^2-\tilde{m}_g^2}
\frac{1}{\left(p_t-k\right)^2-\tilde{m}_n^2}
\frac{1}{\left(p_{\bar t}+k\right)^2-\tilde{m}_m^2}\label{int2}
\eea
where $ {\cal Q} = p_t+p_{\bar t}$ and $\tilde{m}_n \equiv
m_{\tilde{t}_n}$, $m, n = 1, 2$.

The $CP$--violating contribution to $\Im m g_R$ from the
gluino--stop--sbottom loop (see Fig.1)
reads:
\beq
\frac{\Im m g_R}{m_W}=-\frac{\alpha_s}{3\pi}\tilde{m}_g\sin
2\tilde{\theta}\sin
(\phi_A-\phi_{\tilde{g}})\left[I_{1}-I_{2}-2\left(I_{1}^\prime
 -I_{2}^\prime \right)\right]\label{gR}
 \eeq

  Here $I_n$ and $I_n^\prime$ are obtained from (\ref{int1}) and
  (\ref{int2}) by replacing $p_{\bar t}\rightarrow -p_b$
 and $\tilde{m}_m\rightarrow
m_{\tilde{b}_L}$, $m_{\tilde{b}_L}$ being the $\tilde{b}_L$ mass.
In obtaining (\ref{gR}) we have neglected the mixing of the
 scalar--bottom eigenstates.

We have studied the dependence of $\Im m d^\gamma$, $\Im m d^Z$,
$\Im m g_R$,  $\xi_N^{t,\tilde{t}}$ and the CP violating quantities $O_i$
and $M_i$ on $\sqrt{s}$, the top--quark mass $m_t$, and the
SUSY parameters $m_{\tilde{g}}$, $m_{\tilde{t}_1}$ and
$m_{\tilde{b}_L}$. In our numerical analysis we have  assumed
 maximal mixing, $\sin2\tilde\theta = 1$, and maximal $CP$
--violation, $\sin (\phi_A-\phi_{\tilde{g}}) = 1$.
In the presentation of our results we choose
the following set of reference values: $\sqrt{s}=500$~GeV,
$m_t=175$~GeV,  $m_{\tilde{g}}=200$~GeV, $m_{\tilde{t}_1}=150$~GeV,
$m_{\tilde{t}_2}=400$~GeV, and $m_{\tilde{b}_L}=200$~GeV. In the
figures below we show the dependence of $\Im m d^\gamma$, $\Im m d^Z$,
$\Im m g_R$ and the quantities $O_i$
and $M_i$ on one of the parameters, with the other parameters
being fixed at their reference values.

We first show in Figs. 2a, b, and c the dependence of the
 $\Im m d^\gamma$ and $\Im m d^Z$ on the beam energy
$\sqrt{s}$, and on $m_{\tilde{g}}$ and $m_{\tilde{t}_1}$,
respectively. Typically they have values
in the range $10^{-3}-10^{-2}$.

Figs. 3a and b show the dependence of $\Im m g_R$ on the
SUSY parameters $m_{\tilde{g}}$ and $m_{\tilde{b}_L}$.
Roughly, $\Im m g_R$ is an order of  magnitude smaller then the
dipole moments ( $\Im m g_R \approx
10^{-4}-10^{-3}$). As can be seen,
 the dipole moments $\Im m d^{\gamma , Z}$ and $\Im m g_R$ exhibit a
rather strong dependence on
 $m_{\tilde{t}_1}$, changing sign at $m_{\tilde{t}_1}=400$~GeV.
( For our particular set of chosen parameters
$m_{\tilde{t}_1}=400$~GeV corresponds to
$m_{\tilde{t}_1}=m_{\tilde{t}_2}$. At this point eqs. (\ref{dgamma})
and (\ref{gR}) immediately lead to $\Im m d^\gamma =0$ and $\Im m g_R
= 0$; eq. (\ref{dZ}) implies $\Im m d^Z =0$, but only for the considered case
of maximal mixing. )

Figs. 4a, b, c and 5a, b, c exhibit the dependence of the $CP$ -- violating
quantities
$M_i$, and $O_i$, i=1,...,4, on $\sqrt{s}$,
$m_{\tilde{g}}$ and $m_{\tilde{t}_1}$. Note that the observables ${\cal O}_i$
and ${\cal M}_i$ as
defined in eqs. (\ref{calO}) and (\ref{calM}) are ${\cal O}_i = 2 O_i$ and
${\cal M}_i = 2 M_i$. Above the threshold region and up to
$\sqrt{s} \approx 1$~TeV the $M_i$ and the $O_i$ are only weakly dependent on
$s$. $O_2$,
$O_4$ and $M_4$ are the largest. Their values are of the order of
 $10^{-3}$. The
dependence on $m_{\tilde{g}}$ and $m_{\tilde{t}_1}$ reflects the dependence of
$\Im m d^{\gamma,Z} $
and $\Im m g_R$ on these parameters, changing sign at $m_{\tilde {t}_1} =
m_{\tilde
{t}_2} = 400$ ~GeV.

\subsection*{7. Concluding remarks}

Here we have considered $T$--odd triple product correlations and corresponding
 asymmetries for
$t\bar t$ production in $e^+e^-$ annihilation with subsequent semileptonic
 decays of $t$ or $\bar t$.

The contribution to the $T$--odd correlations induced by possible
$CP$--violating interactions
have been studied.  $T$--odd $CP$--violating observables
${\cal O}_i$, eq. (\ref{calO}), and ${\cal M}_i$, eq. (\ref{calM}), have
 been defined. Numerical
predictions within the MSSM have been given, and their dependence on the
 SUSY parameters and
$\sqrt s$ has been presented. Our analysis shows:

1. Above threshold there is only a weak dependence on $\sqrt s$. The quantities
${\cal O}_2$, ${\cal
O}_4$ and ${\cal M}_4$ are larger than the others. Their values at $\sqrt s =
500$~GeV
are of the order of $10^{-3}$ for $m_{\tilde{t}_1} \leq 200$~GeV and $m_{\tilde
g} \lsim 500$~GeV.
Note that the observables ${\cal O}_4$ and ${\cal M}_4$ have also the advantage
that they involve
only the momenta of the $t$ and $\bar t$ decay products.

2. There is a marked dependence on $m_{\tilde{t}_1}$. At
$m_{\tilde{t}_1} = m_{\tilde{t}_2}$ both ${\cal O}_i$ and ${\cal
M}_i$ change sign.

3. The observables  ${\cal O}_i$ and ${\cal M}_i$ considered  exhibit less
sensitivity to $CP$
violation in top decay. This is due to the fact that i)
$\Im m g_R$ is smaller than $d^{\gamma,Z}$, and ii) the
contribution of $\Im m g_R$
to ${\cal O}_i$ and ${\cal M}_i$ is proportional to
the degree of $t(\bar t)$ polarization induced by the SM.

\subsection*{Acknowledgements}

We thank Helmut Eberl for his constructive assistance in the
evaluation of the loop integrals.
E.C.'s  work has been supported by the Bulgarian National
Science Foundation, Grant Ph-16. This work was also supported by the 'Fonds
zur F\"orderung der wissenschaftlichen Forschung' of Austria, project
 no. P10843
-PHY.

\newpage

\subsection*{ Figure Captions}
\begin{enumerate}
\item[{\bf Fig. 1:}] Diagrams inducing CP violation in top production and
decay.

\item[{\bf Fig. 2:}] The imaginary parts of the dipole moments $\Im m d^\gamma$
and $\Im m  d^Z$ as
a function of (a) $\sqrt{s}/2$, (b) $m_{\tilde{g}}$, and (c)
$m_{\tilde{t}_1}$, for $m_t=175$~GeV,
$m_{\tilde{t}_2}=400$~GeV, $m_{\tilde{b}_L}=200$~GeV. We have
taken $\sqrt{s}=500$~GeV in (b) and (c),
$m_{\tilde{g}}=200$~GeV in (a) and (c),
$m_{\tilde{t}_1}=150$~GeV in (a) and (b).

\item[{\bf Fig. 3:}] $\Im m g_R$ as a function of (a) $m_{\tilde{g}}$,
and (b) $m_{\tilde{b}_L}$, for $\sqrt{s}=500$~GeV,
$m_t=175$~GeV, $m_{\tilde{t}_1}=150$~GeV,
$m_{\tilde{t}_2}=400$~GeV, with $m_{\tilde{b}_L}=200$~GeV in (a)
and $m_{\tilde{g}}=200$~GeV in (b).

\item[{\bf Fig. 4:}] The CP violating contribution to the observables $M_i$,
$i=1,..,4$, as a function
of (a) $\sqrt{s}$, (b) $m_{\tilde{g}}$, and (c)
$m_{\tilde{t}_1}$, for $m_t=175$~GeV,
$m_{\tilde{t}_2}=400$~GeV, $m_{\tilde{b}_L}=200$~GeV. We have
taken $\sqrt{s}=500$~GeV in (b) and (c),
$m_{\tilde{g}}=200$~GeV in (a) and (c),
$m_{\tilde{t}_1}=150$~GeV in (a) and (b). $M_1$(full line), $M_2$(long-dashed),
$M_3$(short-dashed), $M_4$(dashed-dotted).

\item[{\bf Fig. 5:}] The CP violating contribution to the asymmetries  $O_i$,
$i=1,..,4$, as a
function of (a) $\sqrt{s}$, (b) $m_{\tilde{g}}$, and (c)
$m_{\tilde{t}_1}$, for $m_t=175$~GeV,
$m_{\tilde{t}_2}=400$~GeV, $m_{\tilde{b}_L}=200$~GeV. We have
taken $\sqrt{s}=500$~GeV in (b) and (c),
$m_{\tilde{g}}=200$~GeV in (a) and (c),
$m_{\tilde{t}_1}=150$~GeV in (a) and (b). $O_1$(full line), $O_2$(long-dashed),
$O_3$(short-dashed), $O_4$(dashed-dotted).
\end{enumerate}

\setlength{\unitlength}{1mm}
\begin{center}
\begin{picture}(110,225)
\put(-18.,135){\mbox{\psfig{file=CPall.eps,height=100mm}}}
\put(-7.,80){\mbox{\psfig{file=CPknot1.eps,height=50mm}}}
\put(63.,80){\mbox{\psfig{file=CPloop1.eps,height=50mm}}}
\put(57.,105.5){\makebox(0,0){\bf \Large{=}}}
\put(-7.,25){\mbox{\psfig{file=CPknot2.eps,height=50mm}}}
\put(63.,25){\mbox{\psfig{file=CPloop2.eps,height=50mm}}}
\put(57.,50.5){\makebox(0,0){\bf \Large{=}}}
\put(55,11){\makebox(0,0)[b]{{\Large \bf Figure 1}}}
\end{picture}

\setlength{\unitlength}{1mm}
\begin{picture}(165,230)(0,0)
\put(17,164){\mbox{\psfig{file=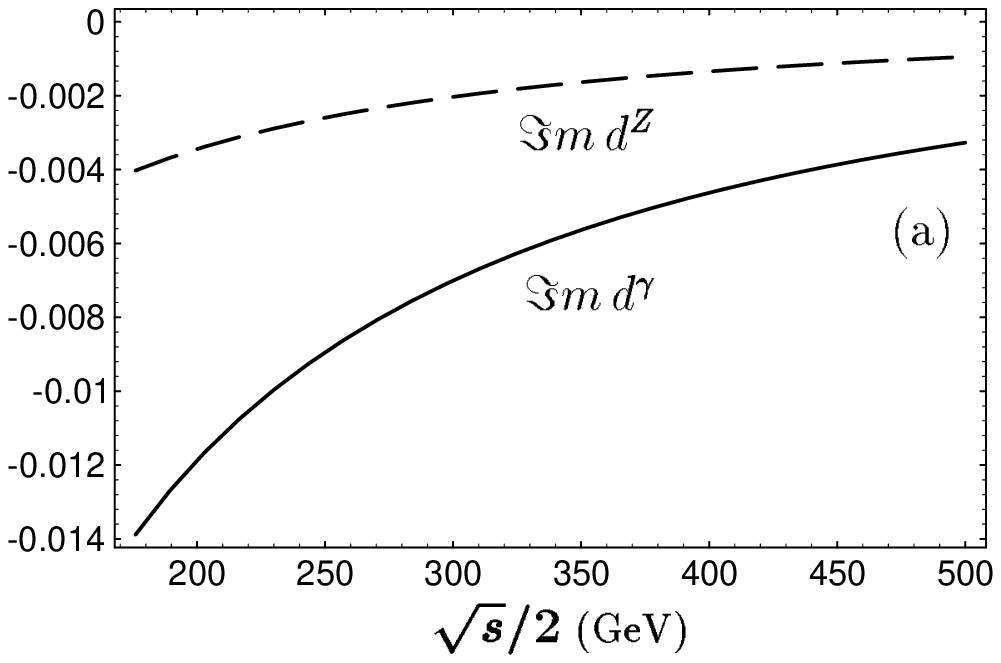,height=6.2cm,width=10.6cm}}}
\put(17,92){\mbox{\psfig{file=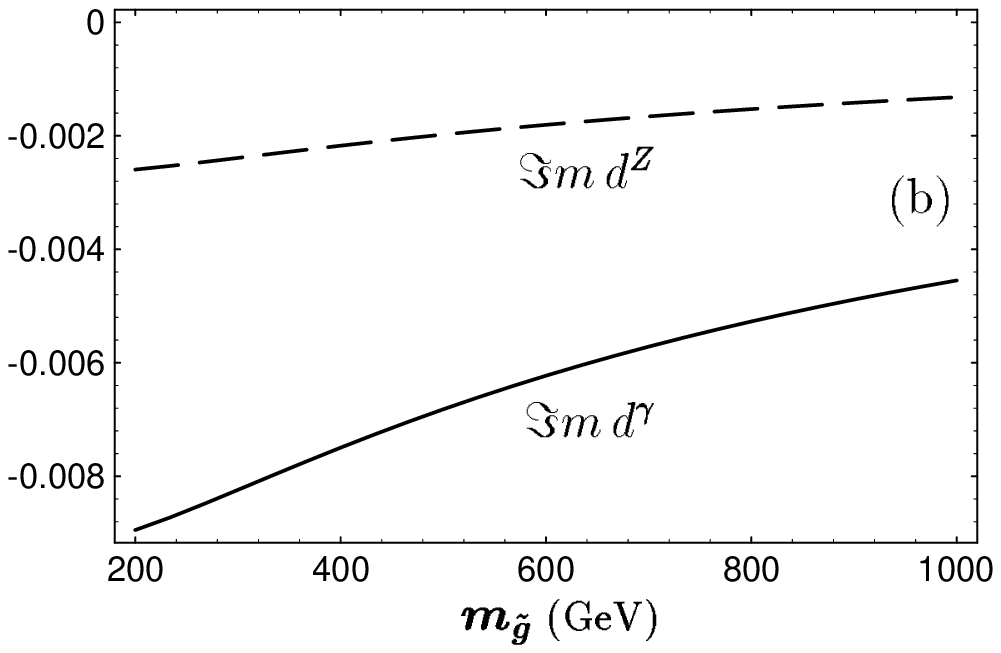,height=6.2cm,width=10.6cm}}}
\put(17,20){\mbox{\psfig{file=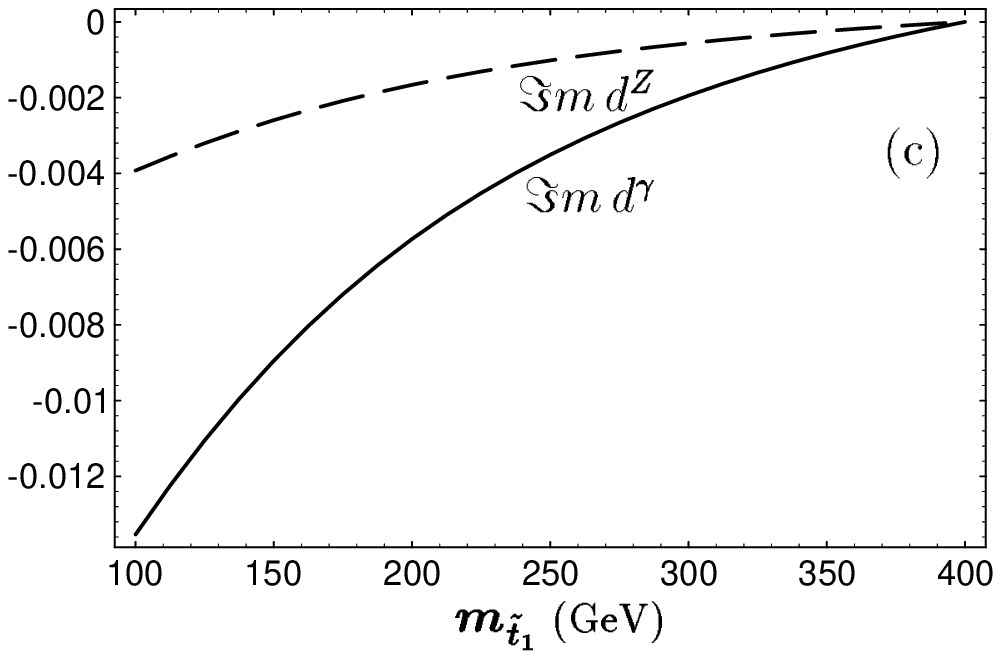,height=6.2cm,width=10.6cm}}}
\put(82.5,10){\makebox(0,0)[b]{{\Large \bf Figure 2}}}
\setlength{\unitlength}{1pt}
\end{picture}

\setlength{\unitlength}{1mm}
\begin{picture}(165,230)(0,0)
\put(30,144){\mbox{\psfig{file=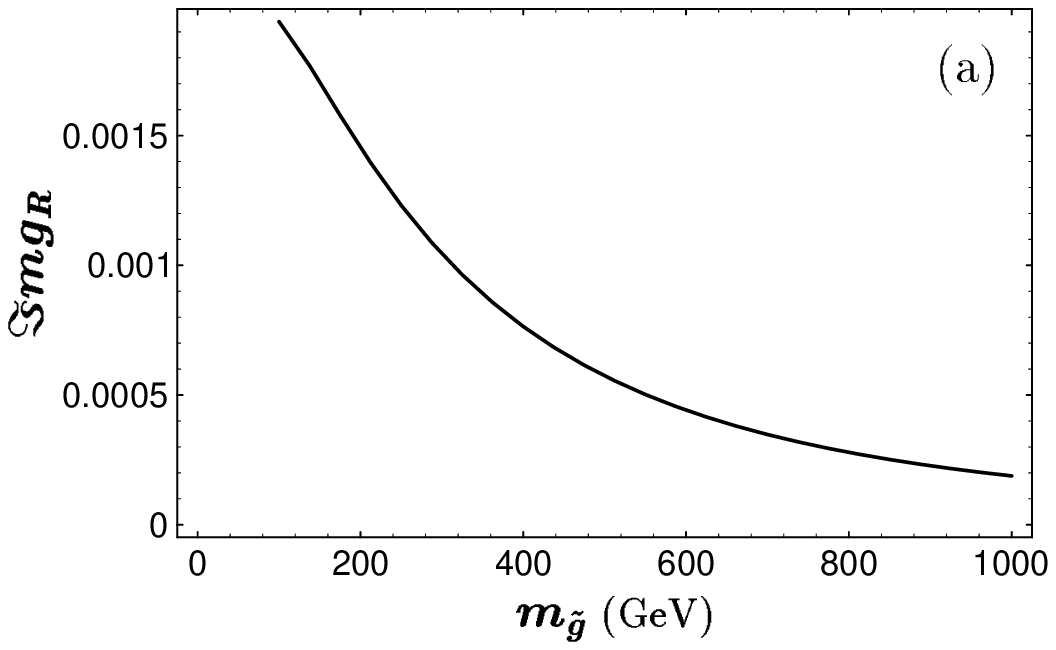,height=6.2cm,width=10.6cm}}}
\put(30,70){\mbox{\psfig{file=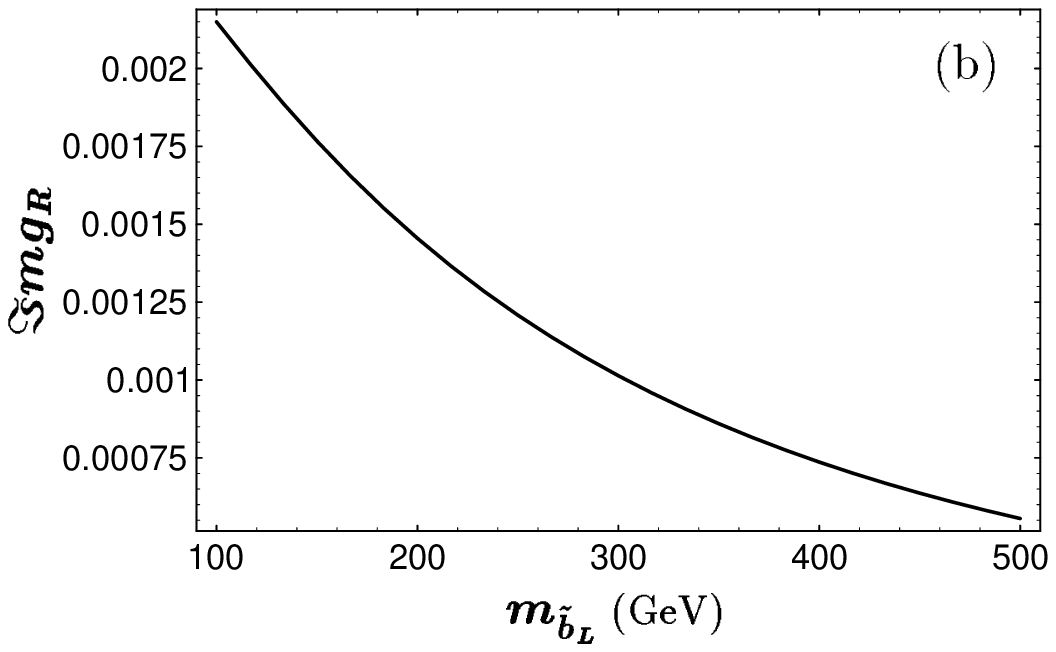,height=6.2cm,width=10.6cm}}}
\put(82.5,60){\makebox(0,0)[b]{{\Large \bf Figure 3}}}
\setlength{\unitlength}{1pt}
\end{picture}

\setlength{\unitlength}{1mm}
\begin{picture}(165,230)(0,0)
\put(16,165){\mbox{\psfig{file=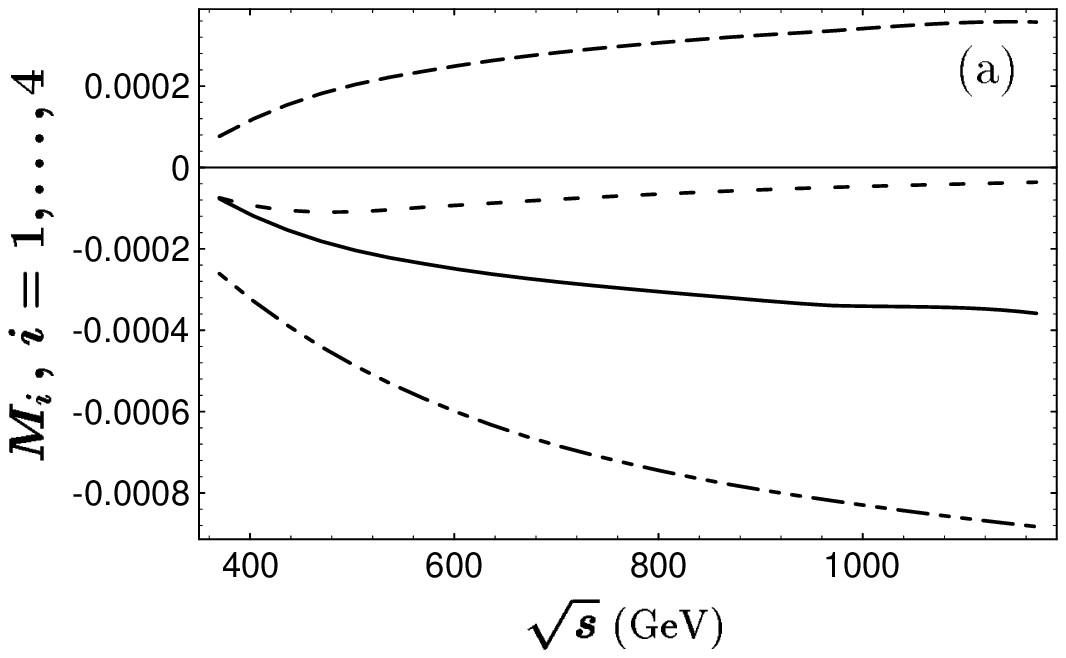,height=6.2cm,width=10.6cm}}}
\put(16,93){\mbox{\psfig{file=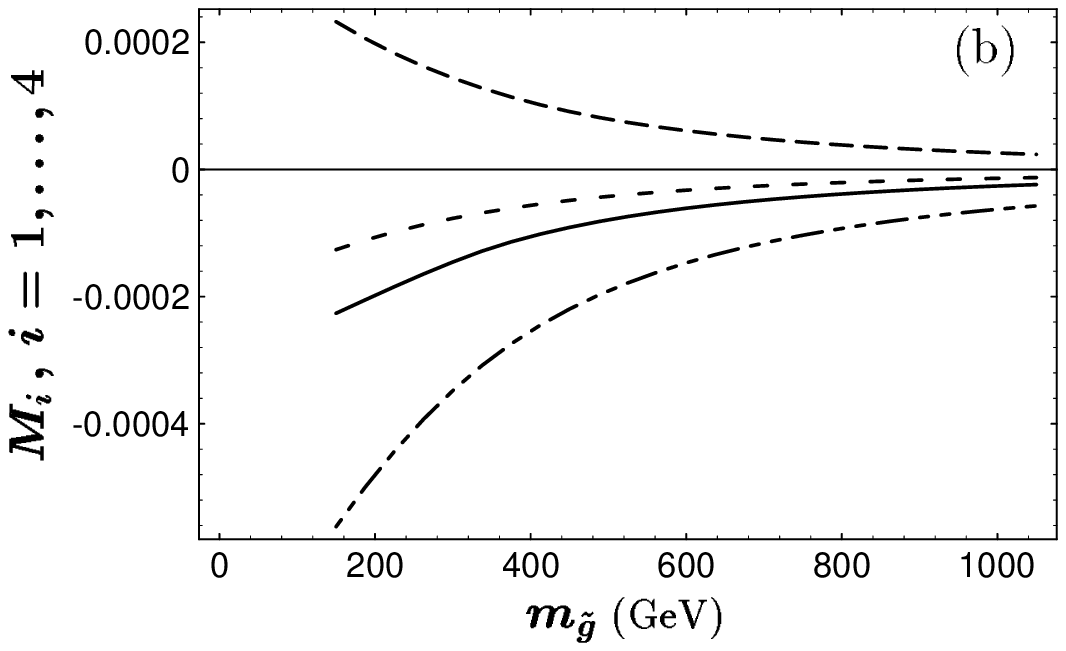,height=6.2cm,width=10.6cm}}}
\put(16,21){\mbox{\psfig{file=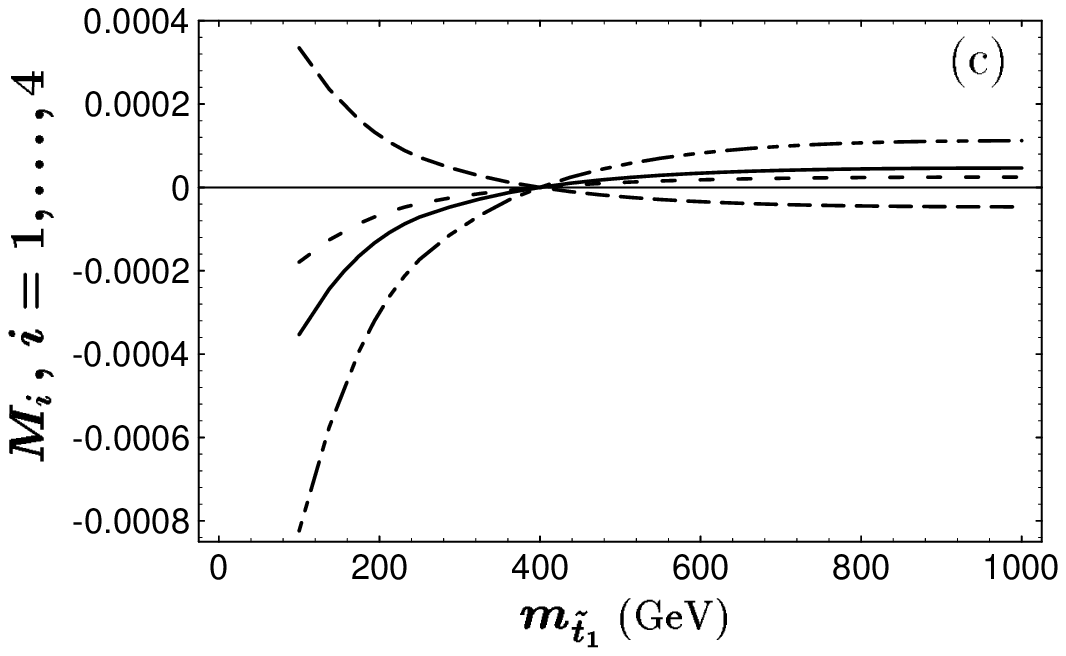,height=6.2cm,width=10.6cm}}}
\put(82.5,11){\makebox(0,0)[b]{{\Large \bf Figure 4}}}
\setlength{\unitlength}{1pt}
\end{picture}

\setlength{\unitlength}{1mm}
\begin{picture}(165,230)(0,0)
\put(20,164){\mbox{\psfig{file=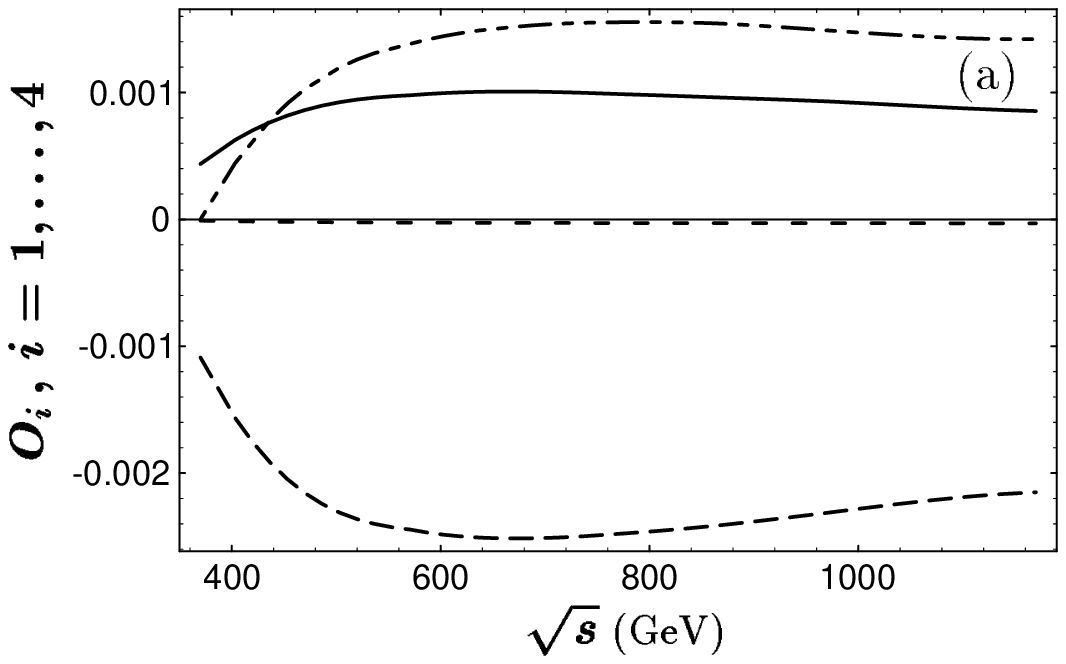,height=6.2cm,width=10.6cm}}}
\put(20,92){\mbox{\psfig{file=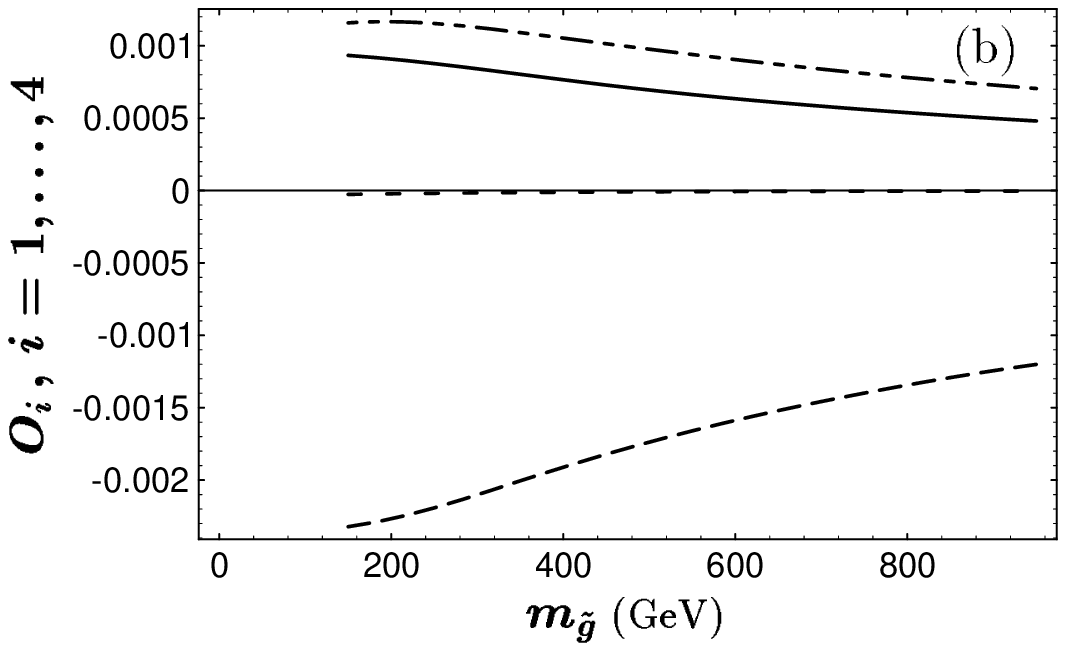,height=6.2cm,width=10.6cm}}}
\put(20,20){\mbox{\psfig{file=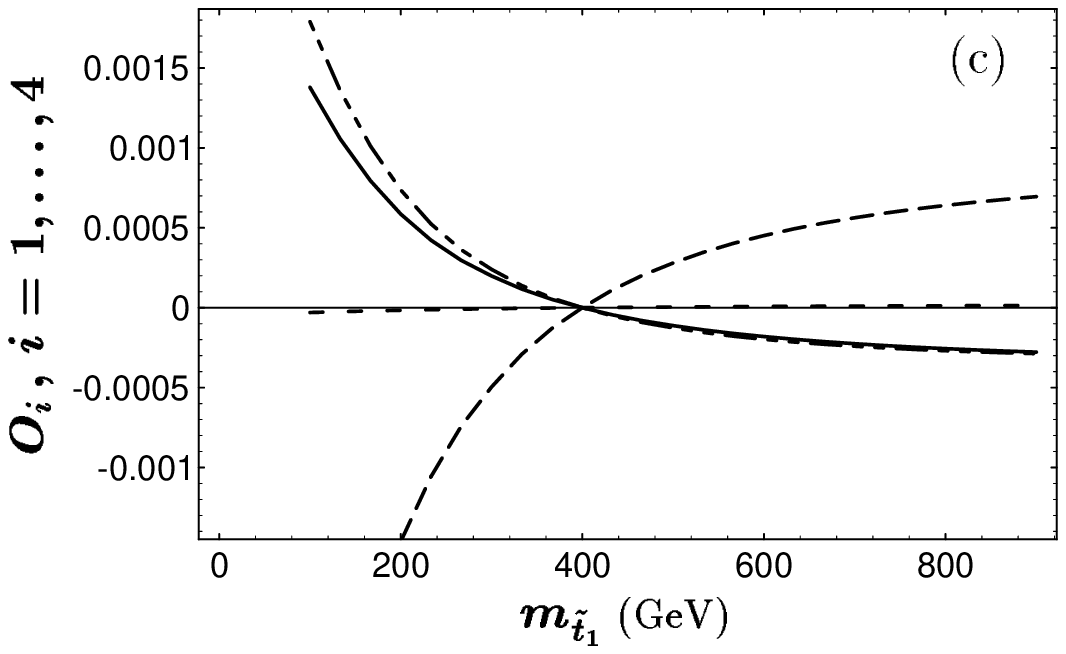,height=6.2cm,width=10.6cm}}}
\put(82.5,10){\makebox(0,0)[b]{{\Large \bf Figure 5}}}
\setlength{\unitlength}{1pt}
\end{picture}
\end{center}

\end{document}